\documentclass{article}
\usepackage{hiph-preprint}
\usepackage{graphicx}
\usepackage{amssymb}
\volnumber{22} \issuenumber{1} \edyear{2005}                             
\frompage{000} \topage{000}                                              
\recrevdate{~}
 \newcommand{\ga}{\alpha}
\newcommand{\gb}{\beta}
\newcommand{\gc}{\gamma}

\newcommand{\gL}{\Lambda}

\def\rightmark{Light front field theory of relativistic quark matter}
\def\leftmark{M.~Beyer, S.~Mattiello, and S.~Strauss}
 
\title{Light front field theory of relativistic quark matter}
\authors{ 
{Michael Beyer$^1$, Stefano ~Mattiello$^1$,  and Stefan Strau\ss$^1$  %
\index{Beyer, M.} 
\index{Strauss, S.} 
\index{Mattiello, S.} 
}\\[2.812mm]
{\normalsize
\hspace*{-8pt}$^1$ Institute of Physics, Universit\"{a}t Rostock, 18051 Rostock, Germany}}
 
\abstract{Light-front quantization to many-particle systems
  of finite temperature and density provides a novel approach
  towards a relativistic description of quark matter and allows us to
  calculate the perturbative as well as the non-perturbative regime of QCD.
  Utilizing a Dyson expansion of light-front many-body Green functions we have
  so far calculated three-quark, quark-quark, and quark-antiquark correlations
  that lead to the chiral phase transition, the formation of hadrons and color
  superconductivity in a hot and/or dense environment. Presently, we use an
  effective zero-range interaction, to compare our results with the more
  traditional instant form approach where applicable.}

\keyword{quark matter, light-front quantisation, finite temperature}
\PACS{11.10.Wx, 
      12.38.Mh 
      25.75    
}
 
\makeindex
\begin{document}
 
\maketitle
\section{Introduction}\label{intro}
Light front quantization recognized by Dirac~\cite{Dirac:cp} provides a
framework to describe the perturbative and the nonperturbative regime of
quantum chromodynamics (QCD), see e.g.~\cite{Brodsky:1997de}. A description
involving both regimes is necessary, if one is interested in the dynamics
driving, e.g., the chiral and the confinement-deconfinement transitions that are
discussed during this conference.
Recently, light front quantization has been successfully 
generalized to finite temperature field theories
\cite{Beyer:2001bc,Alves:2002tx,Weldon:2003uz,Kvinikhidze:2003wc,Das:2003mf,Raufeisen:2004dg}.
It has been shown for the
NJL model that the basic features such as spontaneous symmetry breaking and
chiral symmetry restoration at finite temperature can be
achieved \cite{Beyer:2005rd,Strauss:2004fy}. Within the Dyson expansion
of many-body Green functions light front quantization 
enables us to calculate properties of
three-quark clusters (viz. nucleons) in hot and dense quark
matter \cite{Beyer:2001bc,Beyer:2003ag}. 
This sets up the framework to treat three-quark correlations in quark
  matter and investigate the change from quark matter to nuclear matter.
There is a perspective to use nonzero temperature light-front field theory
to tackle in particular the high density region. Most notable,
the sign problem mentioned in a different context may not occur in this
Hamiltonian approach and one should be able to develop numerical techniques
(similar to the ones used in lattice QCD) to evaluate expectation values of
observables, in particular partition functions.

\section{Light-front statistical physics}
\label{sec:statphys}

A formal framework of covariant calculations at finite temperatures in {\em
  instant} form has been given in Refs.~\cite{Israel:1976tn,isr81,Weldon:aq}
in a different context.  In {\em light-front} quantization the grand canonical
partition function can then be written as
\begin{equation}
Z_G  =\mathrm{Tr}\ \exp\left\{-\frac{1}{T}\left(P_++P_- 
- \mu N\right)\right\}.\label{eqn:part}
\end{equation}
where $T$ is the temperature and $\mu$ the chemical potential, both Lorentz
scalars, and $P^\pm$ and $N$ are defined in light front
quantization~\cite{Brodsky:1997de}. For a grand canonical ensemble in
equilibrium the Fermi distribution functions are given by
\begin{equation}
f^\pm(k^+,\vec k_\perp)=
\left[\exp\left\{\frac{1}{T}\left(\frac{1}{2}k^-_{\mathrm{on}}
+\frac{1}{2}k^+\mp\mu\right)\right\}+1\right]^{-1}
\label{eqn:fermipm}
\end{equation}
where $k^-_\mathrm{on}=(\vec k_\perp^2 +m^2)/k^+$~\cite{Beyer:2001bc}.

\section{Light front cluster many-body Green functions}

The many-body Green functions, can all be defined utilizing light front
quantization.  The causal Green function, e.g., is given by
 \begin{equation}
i{\cal G}_{\alpha\beta}(x^+-y^+)=\theta(x^+-y^+)\langle A_\ga(x^+)
          A_\gb^\dagger(y^+)\rangle 
\mp \theta(y^+-x^+) \langle
          A_\gb^\dagger(y^+) A_\ga(x^+)\rangle\label{eqn:GA}
\end{equation}
The average $\langle\cdots\rangle$ is taken over the exact ground state.  The
Heisenberg operators $A_\ga(x^+)=e^{iH_{\mathrm{eff}}x^+}A_\ga
e^{-iH_{\mathrm{eff}}x^+}$ with $H_{\mathrm{eff}}=P_++P_- -\mu N$, could be
build out of any number of field operators (fermions and/or bosons).  In
equilibrium only one Green function is independent and one may alternatively
use the thermodynamic (or imaginary-time) Green function.  Dyson equations to
disentangle the hierarchy of Green function equations can be established for
both forms, cf.~\cite{duk98} for the nonrelativistic case. In the light-front
formalism the Dyson equations are given by~\cite{Beyer:2001bc}
\begin{eqnarray} 
    i\frac{\partial}{\partial{x^+}}\; {\cal G}_{\ga\gb}({x^+}-{y^+})
     &=&\delta({x^+}-{y^+}) \langle [A_\ga,A_\gb^\dagger]_\pm(x^+)\rangle\nonumber\\
&&    + \sum_{\gc}\int d{\bar x^+}\; {{\cal M}_{\ga\gc}({x^+}- {\bar x^+})}
    \;{\cal G}_{\gc\gb}({\bar x^+}-{y^+}).
\label{eqn:dyson}
\end{eqnarray}
The mass matrix that appears in (\ref{eqn:dyson}) is given by
\begin{eqnarray} 
        {\cal M}_{\ga\gb}({x^+}-{y^+})
        &=& \delta({x^+}-{y^+}) {\cal M}_{0,\ga\gb}({x^+})
        + {\cal M}_{r,\ga\gb}({x^+}-{y^+})\label{eqn:mass}\\
       ( {\cal M}_0{\cal N})_{\ga\gb}(x^+)&=
        & \langle [[A_\ga,H]({x^+}),A_\gb^\dagger({x^+})]_\pm\rangle
\label{eqn:massstat}\\
        ({\cal M}_r{\cal N})_{\alpha\beta}({x^+}-{y^+})&=&
       \sum_\gamma\langle T_{x^+} [A_\alpha,H]({x^+}),
       [A^\dagger_\gb,H]({y^+})]\rangle_{{\rm irreducible}}
\end{eqnarray} 
where ${\cal N}_{\ga\gb}(x^+)=\langle[A_\ga,A_\gb^\dagger]_\pm({x^+})\rangle$.
The first term in (\ref{eqn:mass}) is instantaneous and related to the mean
field approximation, the second term is the retardation or memory term.  We
first solve the mean field problem (neglecting memory).  We use a zero range
interaction, which can be considered as the lowest order effective interaction
appearing in a $1/N_c$ expansion of QCD.  With the help of (\ref{eqn:dyson})
it is possible to derive relativistic in-medium few-body equation to describe
the formation and properties of clusters, viz. nucleons as
three-quark states~\cite{Beyer:2001bc}.

\section{Results}

For the in-medium quark mass we use values of $m(\mu,T)$ given
in~\cite{Beyer:2005rd}. We take $m=336$ MeV for the isolated constituent mass
and determine the coupling strength and the cut-off $\Lambda$ to reproduce the
nucleon mass $m_N=938$ MeV.
\begin{figure}
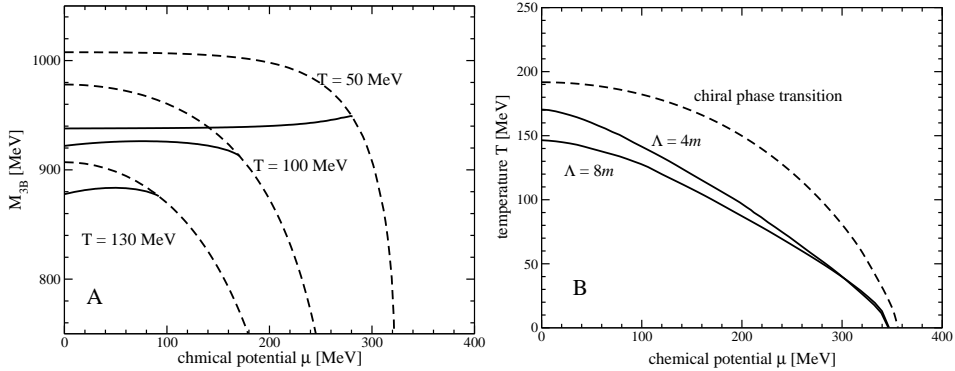

  \includegraphics[width=0.5\textwidth]{M3muLB08}
  \includegraphics[width=0.48\textwidth]{Mott3B}
\caption{\label{fig:Mott}
  {\bf A:} Binding energy of the three-quark bound state (solid) for different
  temperatures as a function of the chemical potential. The respective
  continuum lines are indicated as dashed line, $\gL=8m$. {\bf B:} Mott
  line for the three-body system at rest in the medium for $\gL=4m$ (with
  di-quark substate) and $\gL=8m$ (Borromean nucleon) as extreme cases.}
\end{figure}
The three-body mass $M_3(\mu,T)$ as a function of the chemical potential for
different temperatures and a cut-off parameter $\gL=8m$ are shown in Figure
\ref{fig:Mott}. The chemical potential where the solid lines intersects the
continuum define the Mott transition. The condition is given by
$M_3(\mu_\mathrm{Mott},T_\mathrm{Mott})=3m(\mu_\mathrm{Mott},T_\mathrm{Mott})$
for the three-quark continuum and $M_3(\mu_\mathrm{Mott},T_\mathrm{Mott})=
m(\mu_\mathrm{Mott},T_\mathrm{Mott})+M_2(\mu_\mathrm{Mott},T_\mathrm{Mott})$
for the quark-diquark continuum.
The corresponding Mott
lines are shown in Figure~\ref{fig:Mott} along with the chiral phase
transition line calculated in the NJL model.

\section{Conclusion}
We have shown that light front quantization is applicable to finite
temperature systems and leads to meaningful results. Future challenges are
\begin{itemize}
\item Include three-quark correlations in the distribution function, and the
  thermodynamical potentials to see how quark matter actually changes to
  nucleonic matter when the temperature is lowered.
\item The parameter values chosen are fixed to the nucleon mass which
  eventually should give the strength for the two-quark correlation that is
  important in the studies of color superconductivity.
\item The influence of three-quark correlations has so far not been
  investigated in the context of color superconductivity. The Dyson approach
  provides the necessary tools to do so.
\item An exciting prospect of this approach is to directly use
  light-front QCD~\cite{Brodsky:1997de} to evaluate the partition function given in
  (\ref{eqn:part}). This would surely need quite some effort, however, might
  lack the sign problem present in other approaches due to the Hamiltonian
  formulation.
\end{itemize}

  It is a great pleasure to thank the organizers for the fruitful and exciting
  meeting. Work is supported by the Deutsche Forschungsgemeinschaft (DFG).

\vfill\eject

\begin{thebibliography}{99}  
%
\bibitem{Dirac:cp} P.A. Dirac 1949,
Rev.\ Mod.\ Phys.\  {\bf 21} 392

\bibitem{Brodsky:1997de}
S.~J.~Brodsky, H.~C.~Pauli and S.~S.~Pinsky,
Phys.\ Rept.\  {\bf 301}, 299 (1998), and refs. therein.
%
\bibitem{Beyer:2001bc}
  M.~Beyer, S.~Mattiello, T.~Frederico and H.~J.~Weber,
  Phys.\ Lett.\ B {\bf 521} (2001) 33.
%
\bibitem{Alves:2002tx}
V.~S.~Alves, A.~Das and S.~Perez,
Phys.\ Rev.\ D {\bf 66}, 125008 (2002).

\bibitem{Weldon:2003uz}
H.~A.~Weldon,
Phys.\ Rev.\ D {\bf 67}, 085027 (2003).
\bibitem{Kvinikhidze:2003wc}
A.~N.~Kvinikhidze and B.~Blankleider,
hep-th/0305115.
\bibitem{Das:2003mf}
A.~Das and X.~x.~Zhou,
Phys.\ Rev.\ D {\bf 68}, 065017 (2003).

%
\bibitem{Raufeisen:2004dg}
  J.~Raufeisen and S.~J.~Brodsky,
  Phys.\ Rev.\ D {\bf 70} (2004) 085017
\bibitem{Beyer:2005rd}
  M.~Beyer, S.~Mattiello, T.~Frederico and H.~J.~Weber,
  J.\ Phys.\ G {\bf 31} (2005) 21.
%
\bibitem{Strauss:2004fy}
  S.~Strauss, M.~Beyer and S.~Mattiello,
  Few Body Syst. {\bf 36} (2005) 231.
%
\bibitem{Beyer:2003ag}
  M.~Beyer, S.~Mattiello, T.~Frederico and H.~J.~Weber,
  Few Body Syst.\  {\bf 33} (2003) 89

\bibitem{Israel:1976tn}
W.~Israel,
Annals Phys.\  {\bf 100}, 310 (1976).



\bibitem{isr81} W.~Israel, Physics
  {\bf 106A}, 204 (1981).

\bibitem{Weldon:aq}
H.~A.~Weldon,
Phys.\ Rev.\ D {\bf 26}, 1394 (1982).
\bibitem{duk98} 
        Dukelsky J., R\"opke G. and Schuck P.  (1998),
        Nucl. Phys. {\bf A 628} 17-40.
\end{thebibliography}
\end{document}